\documentstyle[aps,epsf]{revtex}
\tightenlines

\newcommand{\be}{\begin{equation}}
\newcommand{\ee}{\end{equation}}
\newcommand{\bea}{\begin{eqnarray}}
\newcommand{\eea}{\end{eqnarray}}
\newcommand{\bean}{\begin{eqnarray*}}
\newcommand{\eean}{\end{eqnarray*}}

\newcommand{\AmS}{{\protect\the\textfont2
  A\kern-.1667em\lower.5ex\hbox{M}\kern-.125emS}}

\begin{document}

\title{Relativistic Bottomonium Spectrum from Anisotropic Lattices}
\author{ X.~Liao and T.~Manke }
\address{Physics Department, Columbia University, New York, NY 10027, USA}
\date{\today}

\maketitle
 
\begin{abstract}
We report on a first relativistic calculation of the 
quenched bottomonium spectrum from anisotropic lattices. 
Using a very fine discretisation in the temporal direction 
we were able to go beyond the non-relativistic approximation
and perform a continuum extrapolation of our results from
five different lattice spacings ($0.04-0.17$ fm) and two anisotropies 
(4 and 5).
We investigate several systematic errors within the quenched 
approximation and compare our results with those from non-relativistic 
simulations.

\end{abstract}


\section{INTRODUCTION}

The recent experimental activities at B-factories have triggered many
theoretical attempts to understand heavy quark phenomenology from
first principles.
The non-perturbative study of heavy quark systems is complicated
by the large separation of momentum scales which are difficult to
accommodate on conventional isotropic lattices.
These problems are particularly severe in heavy quarkonia, where the small
quark velocity separates the high momentum scales around
the heavy quark mass from their small kinetic energy; $m_q \gg m_q v^2$.
While this motivates the adiabatic approximation and potential models,
a full dynamical treatment poses significant computational problems.
The limitations of present computer resources force us to tackle
this problem in some modified framework.
Thankfully there is also a wealth of spectroscopic data against 
which different lattice methodologies and improvement schemes can be tested.  
Once experimental results can be understood with some accuracy
the lattice will provide a powerful tool for other non-perturbative 
predictions such as decay rates and form factors involving heavy quarks.

Several approximations to relativistic QCD have been proposed 
to describe accurately the low energetic phenomenology of 
heavy quarkonia \cite{nrqcd,fermilab}. 
However, the \mbox{(non-)}perturbative control of systematic
errors in those approximations is very difficult
and in practise one has to rely on additional approximations. 
The high precision results for the spin structure in 
non-relativistic bottomonium calculations\cite{nrqcd_omv4} 
have been hampered by large systematic errors which are difficult to control within this effective theory.
Higher order radiative and relativistic corrections are sizeable for 
bottomonium \cite{omv6,spitz} and even more so for charmonium 
\cite{trottier}. 
Still more cumbersome are large observed scaling violations 
\cite{trottier,nrqcd_nf2} that cannot be controlled by taking the continuum limit.

We take this as our motivation to study heavy quarkonia on anisotropic 
lattices in a fully relativistic framework. This approach has previously 
been used to calculate the charmonium spectrum with unprecedented accuracy 
\cite{tim,ping,lat00,okamoto}. 
Here we extent the method to even heavier quarks and focus on our new results
for bottomonium. Preliminary results were already reported in \cite{lat00}.

There are additional advantages in using anisotropic lattices which
have been employed previously. It has been demonstrated by several authors
that a fine resolution in the temporal direction is useful in 
controlling and extracting higher excitations and momenta 
in the QCD spectrum \cite{morning,hybrid_prl,kuti_prl,collins}.
It has been noticed long ago that anisotropic lattices are also the
natural framework to study QCD at finite temperature \cite{karsch}.
This has been used in \cite{qcd-taro,namekawa} to retain more 
Matsubara frequencies at high temperature. 
Furthermore, it has been suggested that anisotropic lattices 
can circumvent problems arising due to unphysical branches 
of the dispersion relation on highly improved lattices \cite{alford}.

In Section \ref{sec:actions} we introduce the anisotropic gauge and 
quark action for our study. In Section \ref{sec:simulation} we give the
details of our simulation and study systematic errors in Section \ref{sec:systematic}. Our results are discussed in Section \ref{sec:discuss} and
Section \ref{sec:conclusion} concludes this paper.

\section{Anisotropic Lattice Action}
\label{sec:actions}
The basic idea of our approach is to control large lattice spacing artifacts 
from the heavy quark mass by adjusting the temporal lattice spacing, $a_t$, 
such that $m_q a_t < 1$. For the bottomonium system this implies 
$a_t^{-1} > 5$ GeV. Discretisation errors from spatial momenta 
are controlled by $ |{\bf p}| a_s < 1$, which can be satisfied more easily at 
conventional spatial lattice spacings $a_s^{-1} > 1.5$ GeV. 
We retain the isotropy in all the spatial directions.
The continuum limit can then be taken at fixed anisotropy, $\xi = a_s/a_t$.

More specifically, we employ an anisotropic gluon action, which is
accurate up to ${\cal O}(a_s^2,a_t^2)$ discretisation errors:

\bea
S = - \beta \left( \sum_{x, {\rm i > j}} \xi_0^{-1} P_{\rm i j}(x) +
    \sum_{x, {\rm i}}     \xi_0      P_{\rm i t}(x) \right) ~~.
\label{eq:aniso_glue}
\eea
This is the standard Wilson action written in terms of simple plaquettes,
$P_{\mu\nu}(x)$. Here $(\beta,\xi_0)$ are two bare parameters, which 
determine the spatial lattice spacing, $a_s$, and the renormalised 
anisotropy, $\xi$, of the quenched lattice. 
For the heavy quark propagation in the gluon background 
we used the ``anisotropic clover'' formulation as first described in 
\cite{tim,ping}. The discretised form of the continuum Dirac operator, 
$Q=m_q+D\hskip -0.22cm \slash $, reads
\bea
Q  & = &  m_0 + \nu_t~W_0 \gamma_0 + \nu_s~W_i \gamma_i - \nonumber 
\frac{a_s}{2}\left[ c_t~\sigma_{0k}F_{0k} + c_s~\sigma_{kl}F_{kl} \right]~~, 
\nonumber \\
W_\mu & = & \nabla_\mu - (a_\mu/2) \gamma_\mu \Delta_\mu ~~.
\label{eq:aniso_quark}
\eea
Here $\nabla_\mu$ denotes the symmetric lattice derivative
and $a_\mu^2 \Delta_\mu q(x) \equiv U_\mu(x)~q(x+\mu) -2~q(x) + U_{-\mu}(x)~q(x-\mu)$.
For the electromagnetic field tensor $F_{\mu \nu}$ we choose the traceless cloverleaf
definition which sums the 4 plaquettes centered at point $x$ in the $(\mu, \nu)$ plane:
\be
a_\mu a_\nu~F_{\mu \nu}(x) \equiv \frac{i}{2} \left[ P_{\mu \nu}(x) + P_{\nu \bar \mu}(x) +  P_{\bar \mu  \bar \nu}(x) + P_{\bar \nu \mu}(x) - h.c. \right]~~.
\ee
As has been demonstrated by P. Chen \cite{ping}, 
Eq. (\ref{eq:aniso_quark}) is equivalent to the 
continuum action (up to ${\cal O}(a^2)$ errors)
through $O(3)$ symmetric field redefinition:
\be
q(x) \to \left(1 + \frac{\Omega_m}{2} a_t m_q + \frac{\Omega_t}{2} a_t \slash \hskip -0.2cm \nabla_t + \frac{\Omega_s}{2} a_s \slash \hskip -0.2cm \nabla_s \right) q(x)~~.
\label{eq:field_trafo}
\ee
In a recent paper \cite{aoki}, S. Aoki {\it et al.} observed that
this 5-parameter fermion action does not generate Euclidean $O(4)$ 
covariant quark propagators. While it is true that the fermion
Green's functions derived from our action will contain non-covariant 
terms at ${\cal O}(a)$, these can be removed by undoing the field 
transformation (\ref{eq:field_trafo}) at the end of the calculation.
Rather than working with a more complicated action where additional
parameters have to be tuned for covariance, it would be easier to implement
this change of spinor basis after the expensive inversion of the
fermion matrix.
However, in this work we do not even need such a final transformation
since we do not study spin-1/2 hadrons.
\footnote{We would like to thank N. Christ for clarifying this point}
We have chosen Wilson's combination, $W_\mu$, of first and second derivative 
terms so as to ensure the full projection property and to remove all doublers.
The five parameters in Eq. (\ref{eq:aniso_quark}) are all related
to the quark mass, $m_q$, and the gauge coupling as they appear in the
continuum action. By tuning them appropriately we can remove all 
${\cal O}(a)$ errors and re-establish Euclidean $O(4)$ symmetry.
%for physical observables.
The classical estimates have been given in \cite{ping}:
\bea
m_0   &=& m_q(1 + \frac{1}{2}~a_sm_q) \\
\nu_t &=& \nu_s \frac{1 + \frac{1}{2}~a_s m_q}{1 + \frac{1}{2}~a_t m_q} \\
c_s   &=& \nu_s \label{eq:c_s} \\
c_t   &=& \frac{1}{2}\left(\nu_s + \nu_t \frac{a_t}{a_s}\right)~~\label{eq:c_t}.
\eea
Simple field rescaling enables us to set one of the above coefficients 
at will. For convenience we fix $\nu_s=1$ and adjust $\nu_t$ 
non-perturbatively for the mesons to obey a relativistic dispersion relation 
(~$c({\bf 0})=1$~):
\bea
E^2({\bf p}) &=& E^2({\bf 0}) + c^2({\bf p})~{\bf p}^2 + {\cal O} ({\bf p}^4) \ldots ~~.
\eea
We also chose $m_0$ non-perturbatively, such that the rest energy
of the hadron matches its experimental value
(\mbox{$M(^{3}S_1^{--})$ =} \mbox{9.46 GeV} for bottomonium). 
For the clover coefficients $(c_s,c_t)$ we take their classical estimates
from Eqs. (\ref{eq:c_s}) and (\ref{eq:c_t}) and augment the action
by tadpole improvement:
\bea
U_i &\to& U_i/u_s \\
U_0 &\to& U_0/u_t~~.
\eea
Following Ref. \cite{ping} the resulting (dimensionless) fermion matrix can then be written as
\be
a_t Q   =  u_t \hat{m}_0 + \nu_t~\hat{W}_0 \gamma_0 + \frac{\nu_s}{\xi_0} ~\hat{W}_i \gamma_i - \nonumber 
\frac{1}{2}\left[ \frac{c_t}{u_t u_s^2}~\sigma_{0k}\hat{F}_{0k} + \frac{c_s}{u_s^3 \xi_0}~\sigma_{kl}\hat{F}_{kl} \right]~~, 
\ee
where we arranged the terms in such a way that the temporal Wilson term
is multiplied by only $\nu_t$. To compute the coefficients of the spatial 
terms we used $\xi/\xi_0 = u_t/u_s$, which is accurate within 3\% \cite{ping,tim_gauge}.
Remember that we can still choose $\nu_s$ at will and a 3\% uncertainty 
for $c_s$ is certainly acceptable -- at our lattice spacings a 
different tadpole prescription has a much more pronounced effect on 
this coefficient.
The tadpole coefficients have been determined from the average link in Landau gauge: 
$u_{\mu} = 1/3~\langle {\rm tr}~U_\mu(x) \rangle_{\rm Landau}$. 
For brevity we will refer to this as the {\it Landau} scheme.
Any other choice for $(c_s,c_t)$ will have the same
continuum limit, but with this prescription we expect only small
${\cal O}(\alpha a)$ discretisation errors. 

\section{Simulation Details}
\label{sec:simulation}
For the generation of quenched gauge field configurations
we employ a standard heat-bath algorithm as it is also used for
isotropic lattices. 
The renormalised anisotropy, $\xi$, is related to the bare parameter $\xi_0$ 
through:
\be
\xi = \eta(\xi,\beta)~\xi_0~~.
\ee
A convenient parametrisation for $\eta$ can be motivated by a one-loop
analysis of the renormalised anisotropy:
\be
\eta(\xi,\beta) = 1 + (1-\frac{1}{\xi})~\frac{\hat \eta_1(\xi)}{\beta}~\frac{1-a_1~g^2}{1-a_0~g^2}~~,
\label{eq:renorm_aniso}
\ee
where $\hat \eta_1(\xi) \approx 1.02$ and almost constant for our range
of $\xi$. We use Pad\'e parameters ($a_0=-0.77810$ and $a_1=-0.55055$)
which have been determined by an excellent fit over a wide 
range of lattice data with different $\beta$ and $\xi$ \cite{tim_gauge}.
Based on Eq.\ref{eq:renorm_aniso} we choose the appropriate $\xi_0$ to realize 
$\xi$ close to 4 and 5 for each value of $\beta=6/g^2=5.7-6.5$ .
Depending on the gauge coupling, we measure hadron propagators
every 100-400 sweeps in the update process, 
which is sufficiently long for the lattices to decorrelate.

\begin{figure}[t]
\epsffile{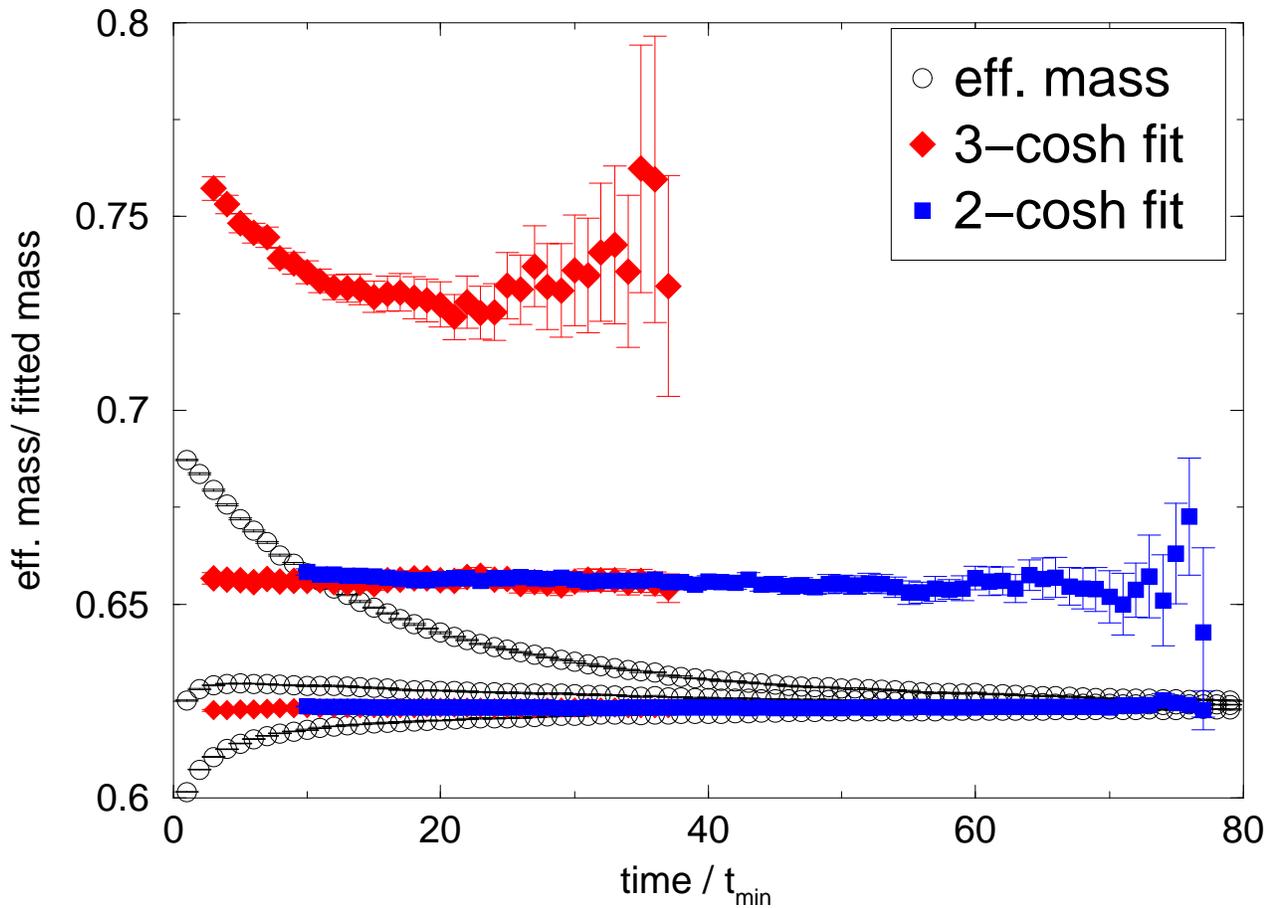}
\caption{Effective masses and fit results. 
We show the effective masses, open circles, for ${^3}S_1$ at 
$(\beta, \xi) = (6.5, 4)$ from 3 different box sources 
with spatial extent of 3, 6 and 9 in lattice units. 
The two larger sources project better onto the ground state as expected. 
The results from the 2-state and 3-state fits are also shown 
as filled squares and diamonds, respectively.
We observe stable fit results for large enough $t_{\rm min}$. 
In this example, we fix $t_{\rm max}=80$.}
\label{fig:effmass}
\end{figure}

The construction of meson operators with given $J^{PC}$ assignment
is standard. As in Refs.\cite{lat00,ukqcd_hybrid}, we use both local and 
extended operators by employing the 16 $\Gamma$-matrices 
and spatial lattice derivatives for the quark-bilinears:
\be
M(x) = \bar q(x) \Gamma_i \Delta_j[U] \Delta_k[U] q(x)~~.
\label{eq:operators}
\ee
To improve the overlap with the ground state we also 
implemented a combination of various iterative smearing prescriptions
for the quark fields and gauge links:

\bea
q(x) &\to& q^{(n)}(x)=(1-\frac{\epsilon}{n} \Delta^2[U])^n q(x) \mbox{\hskip 3.5cm \it Jacobi smearing}\\
U(x) &\to& U^{(m+1)}(x)={\cal P}_{SU(3)}~\left( U^{(m)}(x) + \alpha \sum_{\rm staples} S(x) \right)  \mbox{\it ~~~~~APE smearing}
\eea
For conventional states, we find it 
often sufficient to use (gauge-fixed) box sources of different 
size. We use local sinks throughout.
This setup allows us to extract reliably both the ground state energies and 
their excitations from correlated multi-state fits to several smeared 
correlators, $C^s(t)$, with the same $J^{PC}$:
\be
C^{s}(t, {\bf p}) \equiv \langle M(t,{\bf p})~M^s(0,{\bf p}) \rangle = \sum_{i=1}^{\rm n_{fit}}~a_i^s({\bf p})~\left(e^{-E_i({\bf p})~t} + e^{-E_i({\bf p}) (N_t -t)} \right)~~.
\label{eq:theory}
\ee
Since we are working in a relativistic setting, the second term takes 
into account the backward propagating piece from the temporal boundary.
An example of this is given in Figure \ref{fig:effmass},
where we show the effective masses for ${^3S}_1 = 1^{--}$ 
from three different box sources along with the fit results from
a 2-state and 3-state ansatz.
In this example the smallest box shows clear contributions from 
excited states, while the other two sources project much better onto 
the ground state. We did not further optimize the box size for 
perfect overlap, but since for the largest source the plateau 
is approached from below we can estimate that an ``optimal'' box 
would have a physical extent of 0.2-0.3 fm. This is in good agreement 
with phenomenological expectations about the size of the $\Upsilon$
ground state.
What is more important here for us is to have a variety of different
overlaps which are well suited to constrain the multi-state fits.

\begin{figure}[t]
\epsffile{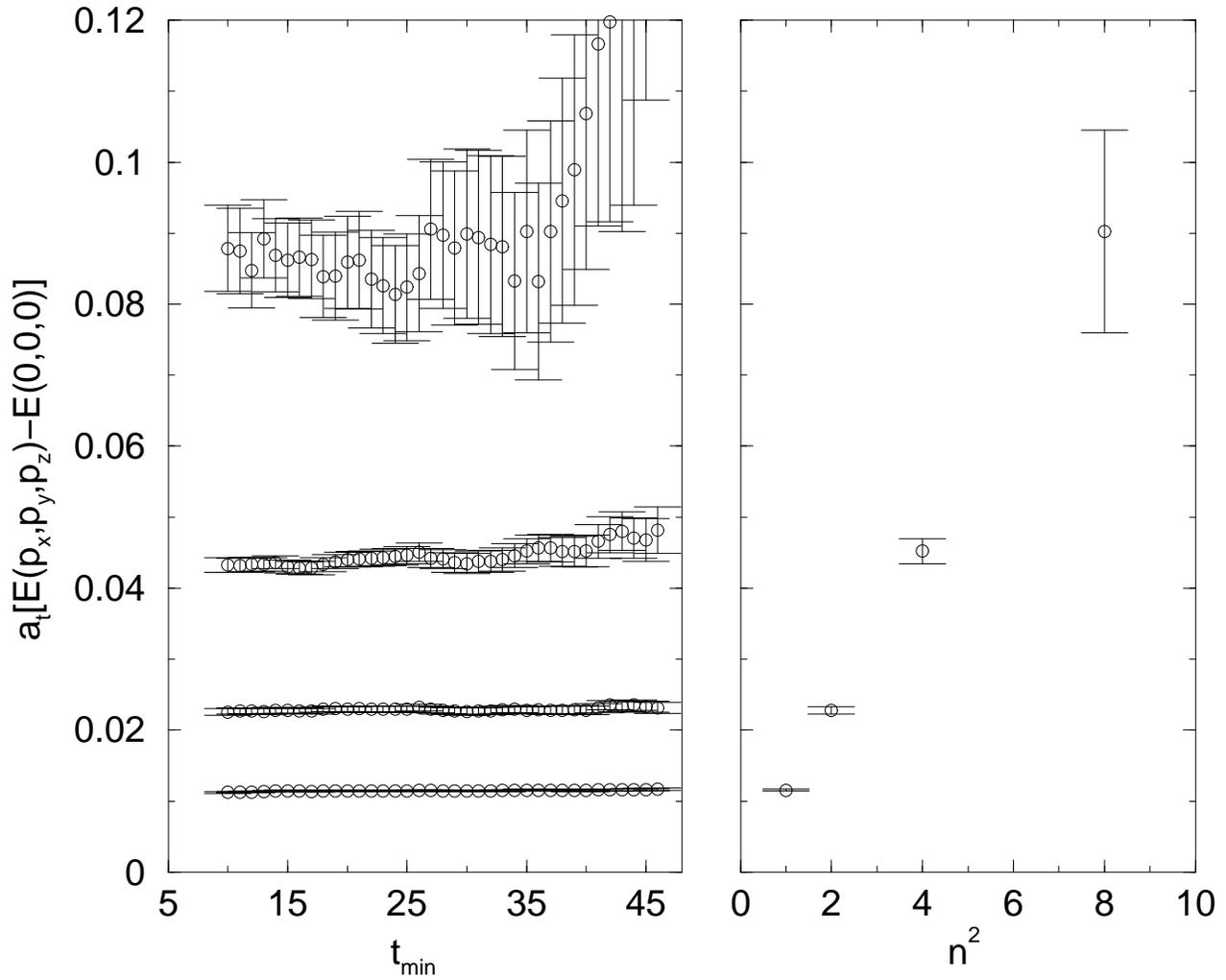}
\caption{Dispersion relation. On the left-hand side we show 
$t_{\rm min}$-plots for the fitted energy difference between  
non-zero momentum states and the ${^3S}_1$ at rest.
On the right hand side we plot the fitted splitting against the
spatial momenta squared, ${\bf p}^2=(2\pi/L)^2~{\bf n}^2$.
Since we tuned $\nu_t$ very carefully, this correspndis to a velocity
of light $c(0)=0.990(11)$.
This example is from $(\beta,\xi)=(5.9,5)$ with  $\nu_t=1.815$.}
\label{fig:dispers}
\end{figure}

To obtain the dispersion relation, $E({\bf p})$, we project the meson
correlator also onto four different non-zero momenta by inserting 
the appropriate phase factors, $\exp{(-i{\bf px})}$, at the sink. 
This is illustrated in Figure \ref{fig:dispers}, where we use
\be
{\bf p} = \left(\frac{2\pi}{N_s}\right)~ {\bf n}  \hskip 1cm  \mbox{with~~~} {\bf n} = (0,0,0),(1,0,0),(1,1,0),(2,0,0),(2,2,0)~~.
\ee

\begin{figure}[t]
\epsffile{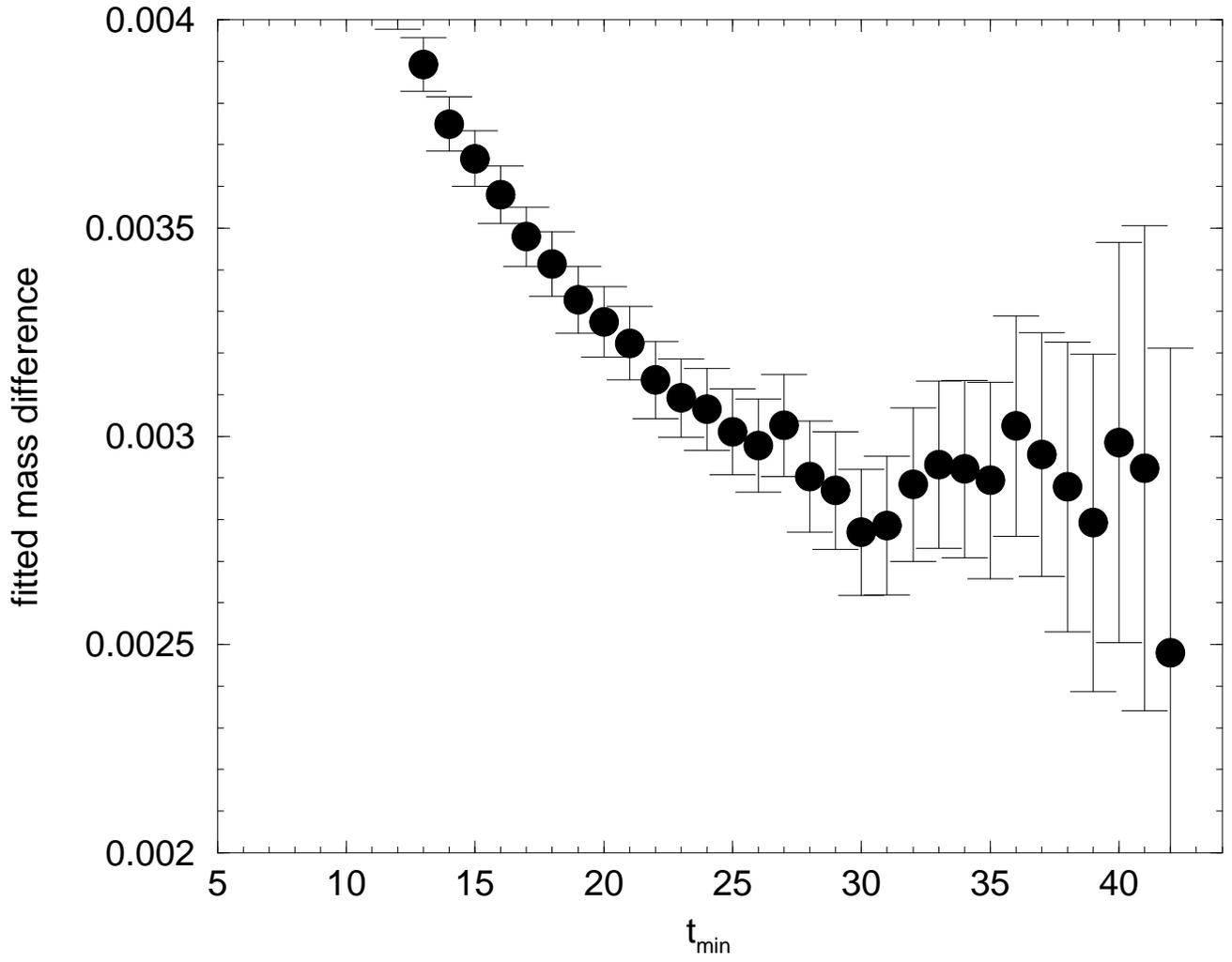}
\caption{Here we show fit results for the ${^3}P_1-{^3P}_0$ splitting
at $(\beta, \xi)=(6.1,4), 16^3 \times 96$. We fix $t_{\rm max}=48$ and vary
$t_{\min}$ in search for a plateau.
Strong correlations between the the two states
allow a very accurate determination of the energy difference.}
\label{fig:3P1-3P0}
\end{figure}

For an estimate of the spin-splittings we take the difference of the
ground state masses from multi-state fits. Because these states are
highly correlated we obtain very accurate estimates for the spin-splittings
compared to the absolute masses. An example of this is shown in 
Figure \ref{fig:3P1-3P0}.
We use jackknife and bootstrap ensembles for our error
estimates and find good agreement in all cases.
In order to call a fit acceptable we require consistency within errors
for different fit ranges and Q-values to be bigger than 0.1.
Our main results are collected in Table \ref{tab:results} along with the
input parameters for the different data sets.

For $(\beta,\xi)=(6.1,4)$ and $(6.3,4)$ we have also analyzed extended 
operators that give access to $2^{++}={^3}P_2$, $D$-states with two
units of orbital angular momentum and exotic hybrid states, 
Table \ref{tab:exotics}. These are more states than previously obtained
from non-relativistic calculations.
Even though our scaling window is not wide enough to make 
a serious attempt of a continuum extrapolation, 
we expect lattice artifacts to be less pronounced for those 
large states. At (6.1,4) we find a good agreement of our results
and those reported in \cite{hybrid_prl,kuti_prl,ukqcd_hybrid}. 
This is not unexpected since the non-relativistic approximation
is known to work better for hybrid states, which are thought to 
live in a very shallow potential \cite{kuti_prl}.
In particular we find the lowest lying hybrid to be the $1^{-+}$.
We are however concerned that these results are more
sensitive to finite volume effects than those for $S$ and $P$ states,
see Sec. \ref{sec:systematic}.
Especially at (6.3,4) we have reason to believe that our lattice 
is not sufficiently big to accommodate such large wavefunctions,
resulting in an inversion of the characteristic level ordering.
A simulation on bigger volumes is necessary to simulate hybrid
states more accurately.

In the following we will focus on the low-lying bottomonium states 
and their spin-structure where statistical errors are small enough 
to allow a critical comparison with results from the 
non-relativistic approximation.

\section{Systematic errors}
\label{sec:systematic}
\begin{figure}[t]
\epsffile{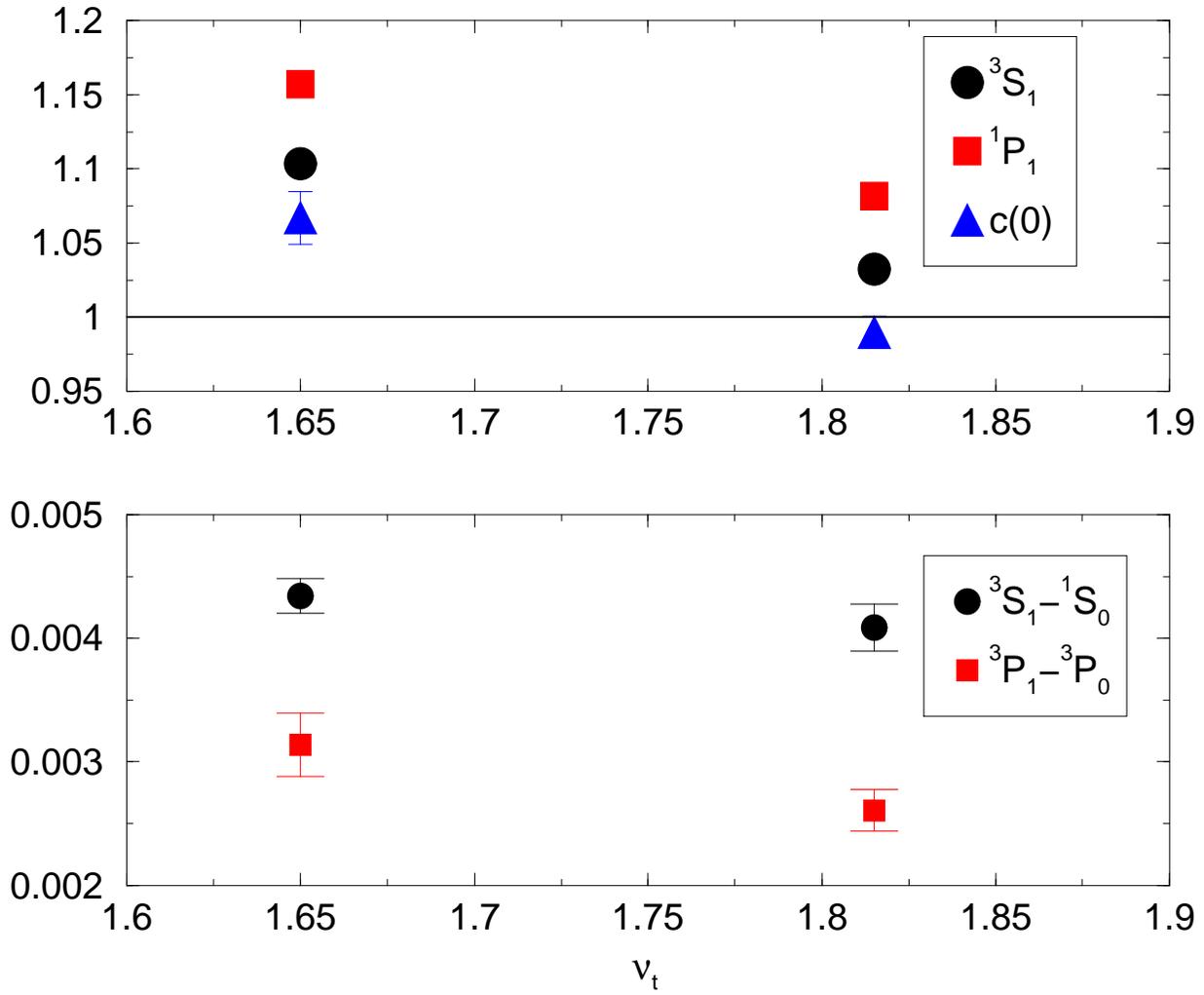}
\caption{$\nu_t$-tuning. We illustrate the non-perturbative tuning of the 
velocity light $(\beta,\xi)=(5.9,5)$. Our best estimate is $\nu_t=1.815$
and we show the effect on spectral qunatities as we change this parameter by 10\%. }
\label{fig:nut_dependence}
\end{figure}

Apart from our main data sets we performed
extensive checks of systematic errors within the quenched approximation.
As mentioned above, we enforce a relativistic dispersion relation by
tuning the bare parameter $\nu_t$ non-perturbatively at each quark mass 
to obtain c=1.0 very accurately within $1-2\%$.
From a classical analysis of the dispersion relation we expect
$c \propto 1/\nu_t$. Indeed, for an arbitrary decrease of $\nu_t$ 
by $10\%$ we observe an increase of the velocity of light by a 
similar amount (7.8(2.1)\%), see Table \ref{tab:nut} and 
Figure \ref{fig:nut_dependence}.
Within the statistical error for the P structure we could not resolve
any significant effect, while the hyperfine splitting showed an 
increase by $8(10)\%$. From those observations, we expect only 
small changes ($<2\%$, $<1$ MeV) due to $c \ne 1$.

\begin{figure}[t]
\epsffile{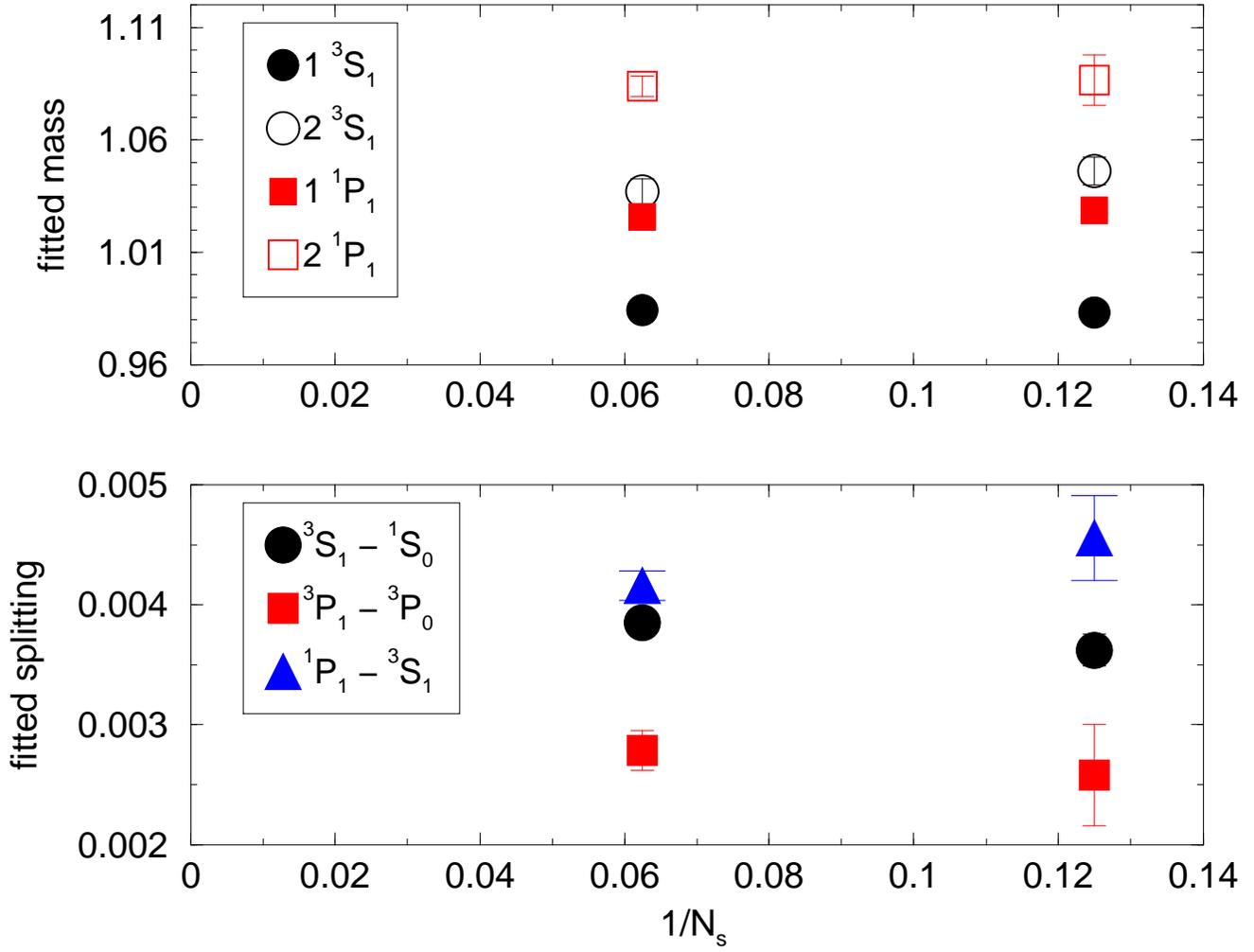}
\caption{Finite volume effects. Here we compare our spectrum results
at $(\beta,\xi)=(6.1,4), N_s^3 \times 96$ for  $N_s=8$ and $16$. }
\label{fig:finite_volume}
\end{figure}

We also tested for finite size effects by explicitly comparing the 
results from two different spatial volumes of $(1.3 \mbox{~fm})^3$ 
and $(0.7 \mbox{~fm})^3$  at $(\beta,\xi) = (6.1,4)$, 
see Figure \ref{fig:finite_volume}.
While changes in the ground state masses amount to only a fraction of 
a percent, the hyperfine splitting shows a slight increase of 6(4)\% and the 
${^1P}_1-{^3S}_1$ splitting is reduced by 10(9)\% when going to the 
larger lattice. Therefore we expect the hyperfine splitting to be
systematically underestimated when $L < 1$ fm. For the fine structure we could 
not resolve any shift within the statistical errors. 
For all but our finest lattice we have $L \approx 1$ fm 
which should be sufficiently big to accommodate the small bottomonium 
ground states in a quenched simulation.

\begin{figure}[t]
\epsffile{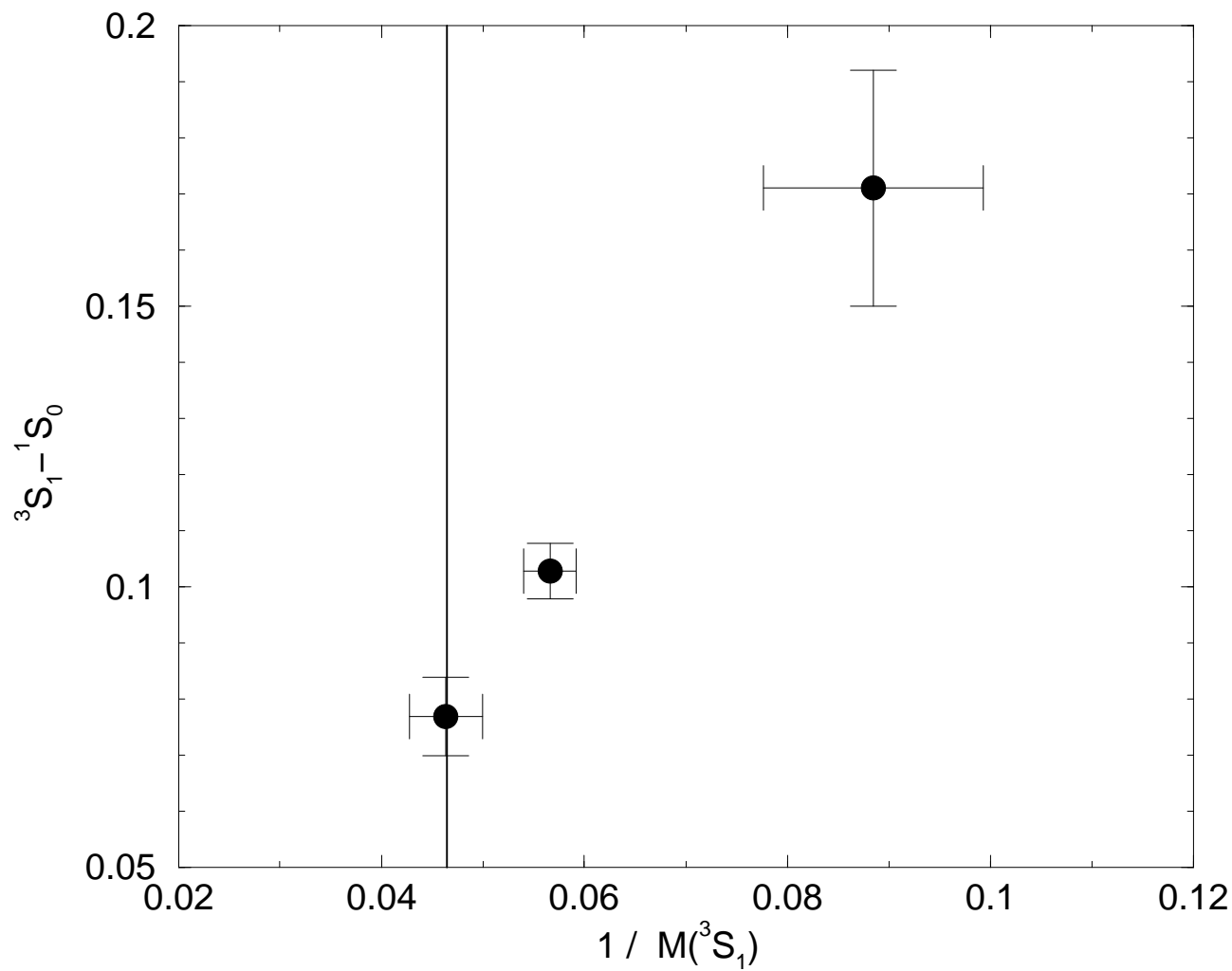}
\caption{Mass dependence. We plot the dimensionless hyperfine splitting against the inverse
of the measured meson mass. The vertical line denotes the Bottomonium mass
at $(\beta,\xi)=(6.1,4)$. The corresponding bare quark masses are $m_0=0.67, 0.40$ and $0.10$.}
\label{fig:mass_dependence}
\end{figure}

Another source of systematic errors is the tuning of the 
bare quark mass parameter, $m_0$.
Especially for the hyperfine splitting we observed a strong 
dependence on the ground state mass, $M(m_0)$, as shown in
Figure \ref{fig:mass_dependence}.
On our coarser lattices we were able to tune the mass very accurately 
to within 10\% of the physical bottomonium mass, but as the tuning 
becomes more and more expensive on our finer lattices, the deviation
from bottomonium reaches 35(15)\% at $(\beta,\xi)=(6.5,4)$. 
We use the potential model prediction, $\Delta E_{\rm hfs} \propto 1/M$,
to rescale all measured $\Delta E_{\rm hfs}$ to the 
physical point, $M=9.46$ GeV.
Our results in Table \ref{tab:results} include these adjustments in both
the central values and their errors.
Since all spin splittings vanish in the static limit we also 
apply the same rescaling technique to the fine structure.

\begin{figure}[t]
\epsffile{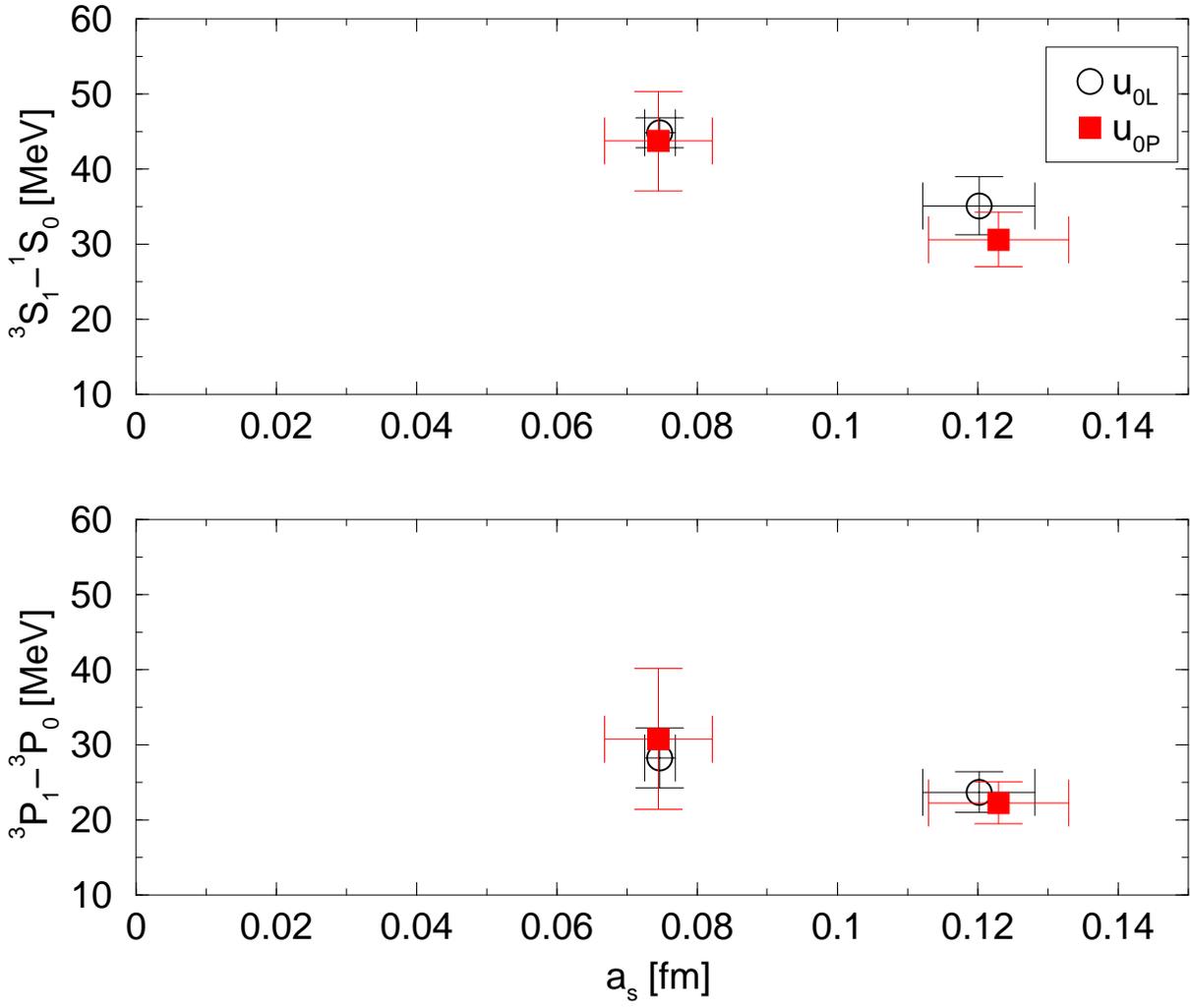}
\caption{Different tadpole prescriptions. Here we compare the effect of
changing the value of $(c_s,c_t)$ by (10\%,2\%). 
Within our statistical errors there is only a minor shift to indicate
the corresponding change in ${\cal O}(\alpha a)$ scaling violations.}
\label{fig:tadpole}
\end{figure}

Finally there will be systematic effects at finite lattice spacing,
which are absent in the continuum limit.
The tree-level estimates for $c_s$ and $c_t$ will experience
a shift due to radiative corrections, thereby changing the magnitude of
the ${\cal O}(\alpha a)$ artifacts.  
In Figure \ref{fig:tadpole} and Table \ref{tab:tadpole} 
we compare the Landau scheme with another popular scheme, 
where the tadpole coefficients are estimated from the plaquette values: 
$3 u_\mu= \langle {\rm tr}~P_{i \mu} \rangle^{1/4}$.
This amounts to a $(10\%, 2\%)$ change in $(c_s,c_t)$ at (5.9, 4).
The corresponding change in the hyperfine splitting is -15(17)\%.
While this may indicate some additional reduction at coarse lattices
(resulting in a slightly worse scaling behaviour),
it is also clear that within our errors we are not very sensitive to
this choice of tadpole coefficients.
Since any remaining difference will vanish in the continuum limit we
did not investigate this dependence further.

\section{Discussion}
\label{sec:discuss}
To convert lattice data into dimensionful quantities we used the 
lattice spacing determined from the ${^1P}_1-{^3S}_1$ splitting which is fixed
to match the experimental value ($\approx 440$ MeV). 
It is well-known that, without dynamical sea quarks in the gluon background, 
the definition of the lattice spacing is ambiguous and one cannot 
reproduce all experimental splittings simultaneously. 
Non-relativistic lattice calculations with two dynamical flavours  
\cite{spitz,nrqcd_nf2} resulted in shifts of up to 5 MeV for the 
hyperfine splitting, but they are also not completely free from ambiguities 
in the lattice spacing and from systematic errors such as 
lattice spacing artifacts, radiative and relativistic corrections.
For the purpose of this paper we accept the shortcomings of the quenched 
approximation and aim to control the other systematic errors instead, 
the combined effect of which could also be as large as 20\% for bottomonium
on the lattices in this study.

\begin{figure}[t]
\epsffile{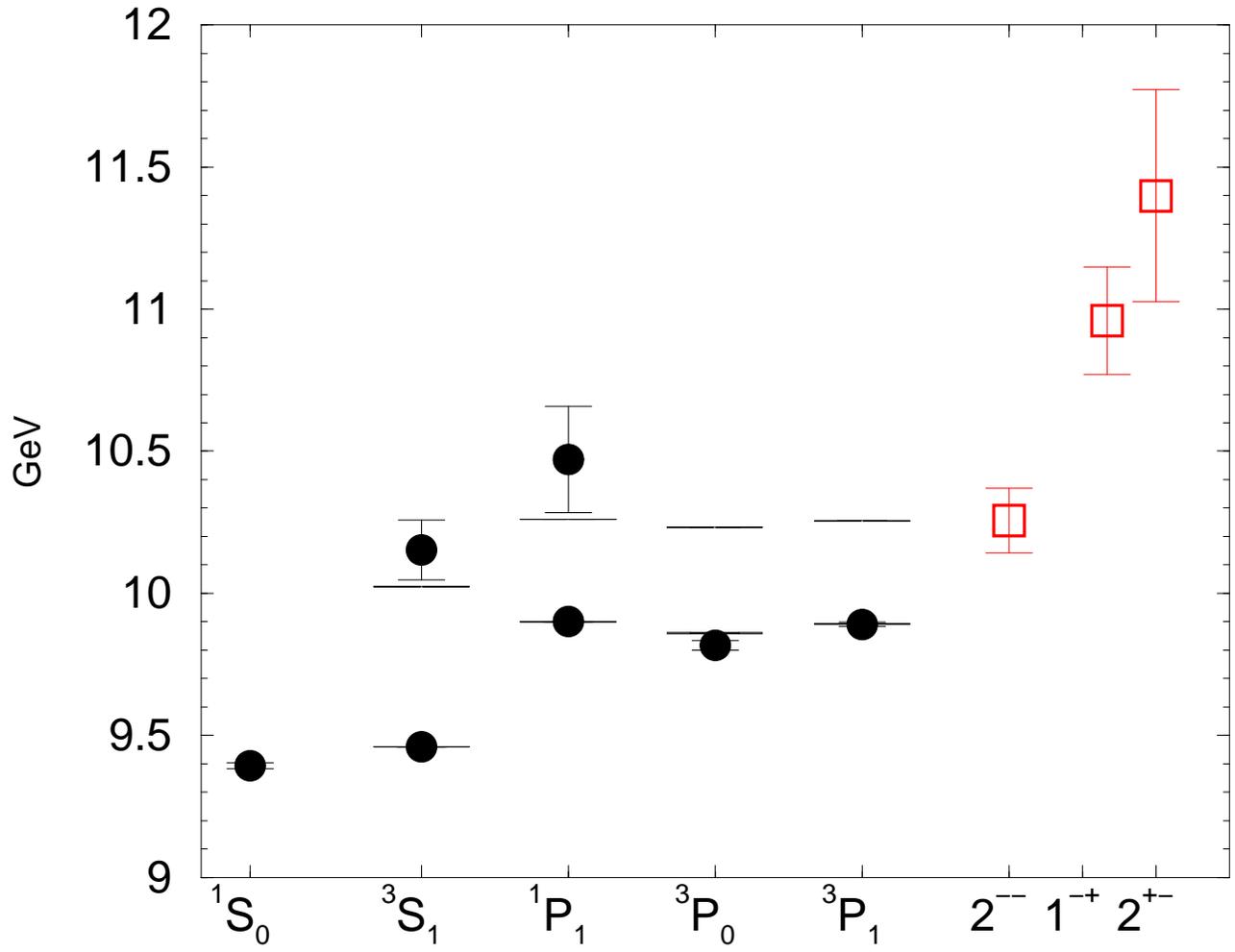}
\vskip +0.5cm
\caption{Relativistic bottomonium spectrum.
We plot our quenched continuum estimates as full circles. 
Selected results for $2^{--}$ and exotic candidates (hybrids) 
from finite lattice spacing, (6.1,4), are shown as open squares. 
The ${^1P}_1-{^3S}_1$ splitting is used to set the scale.
Where available, experimental values are shown as horizontal lines.}
\label{fig:spectrum}
\end{figure}

The main features of the bottomonium spectrum from relativistic lattice
QCD are summarised in Figure \ref{fig:spectrum}, where we can see clearly
the characteristic level ordering as additional angular momentum is
inserted into the ground state. The overall agreement with 
experimental data is impressive, as is the spin-structure predicted 
from first principles. While it is probably too early to investigate 
the spin-splittings in excited states 
($2~{^3S}_1 - 2~{^1}S_0$ has still large statistical errors),
we have rather accurate data for the spin-structure of the ground states.

\begin{figure}[t]
\epsffile{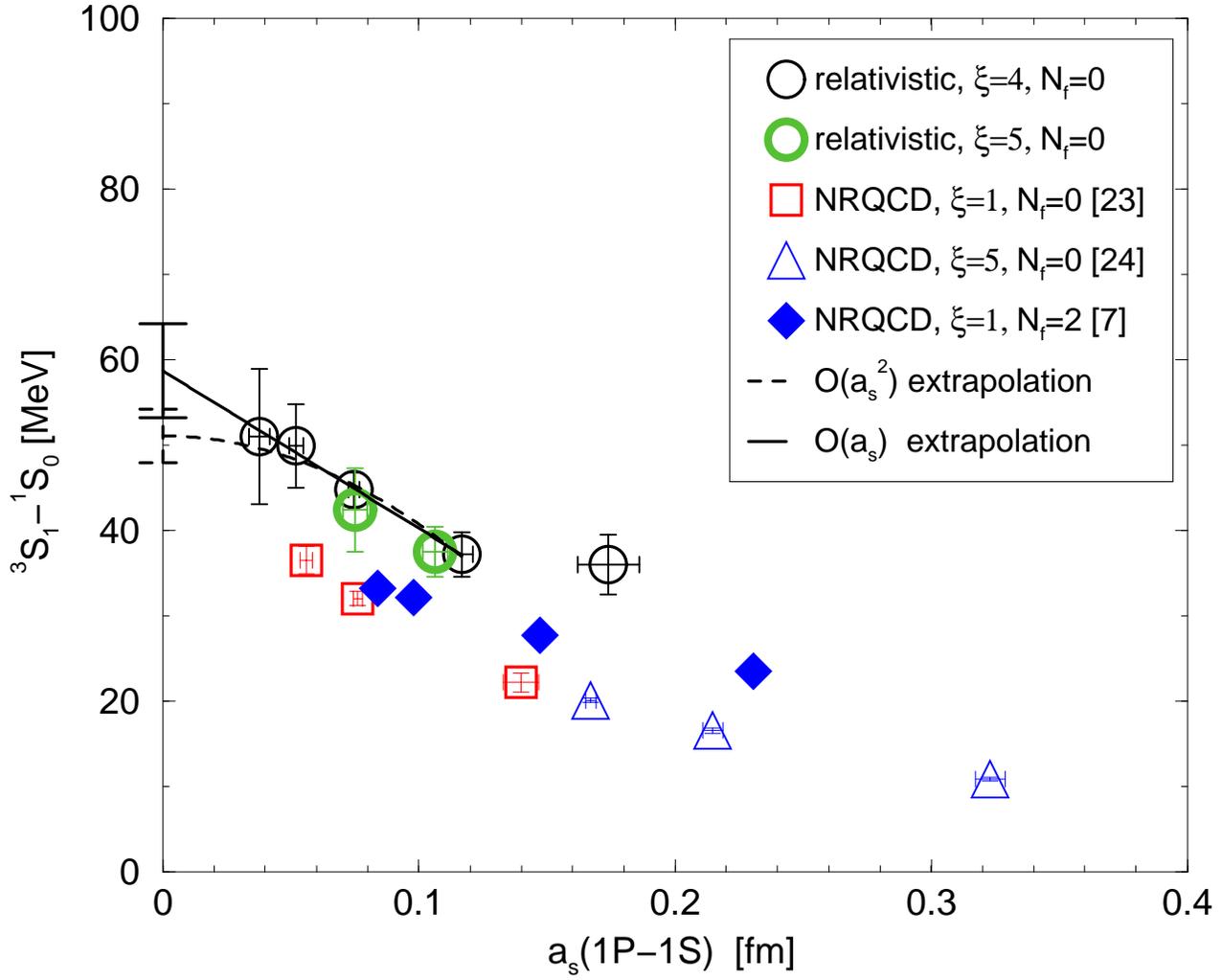}
\caption{Continuum extrapolation of the hyperfine splitting in Bottomonium. Circles denote our
results from different anisotropic lattices. Open squares and triangles are quenched NRQCD results
from isotropic and anisotropic lattices. Filled diamonds are unquenced NRQCD results with two dynamical
flavours at $m_\pi/m_\rho \approx 0.5$. We also show two different continuum extrapolations of our $\xi=4$ 
results, where we assume a simple linear or quadratic form in $a_s$. 
The latter results in a better fit but larger continuum value.}
\label{fig:hfs_vs_as}
\end{figure}

In Figure \ref{fig:hfs_vs_as} we plot the hyperfine splitting against
the spatial lattice spacing at fixed anisotropy and compare it with
previous simulations in this region.
Already at finite lattice spacing we can see significant deviation 
from non-relativistic simulations. There could be several reasons for this.
First, we should expect different discretisation errors from the different
gluonic actions that have been employed in the past. While the unquenched
results in Figure \ref{fig:hfs_vs_as} were obtained from an RG-improved gluon
action with two dynamical flavours \cite{nrqcd_nf2}, many other quenched
simulations where done using the standard plaquette action 
\cite{nrqcd_omv4,omv6,spitz,nrqcd_scaling}.
Non-relativistic results from coarse and anisotropic lattices
also exist \cite{mont99} and they seem to show scaling violations
similar to those from unimproved gluon actions, even though
a plaquette-plus-rectangle (Symanzik) action had been used.
This could be an indication of comparatively large ${\cal O}(\alpha a)$
corrections which can compete with ${\cal O}(a^2)$.

Some of those results mentioned also use a different prescription for 
the tadpole coefficients than in our present work.
We want to stress again that the tadpole prescription is merely an empirical
way of keeping radiative corrections small. It has been suggested
by several authors that the Landau scheme is particularly suitable 
to account for radiative corrections due to tadpole diagrams 
\cite{trottier,landau}.
In the continuum limit ($U_\mu \to 1$) both definitions (Landau and plaquette)
will be identical, but at finite lattice spacing they result 
in different estimates for the clover coefficients. 
We have demonstrated above that a lower value for $u_0$ will result 
in a slightly larger hyperfine splitting. 
A more pronounced effect has been observed in \cite{spitz}.
In the non-relativistic framework one can also quantify this shift which has 
been employed in \cite{nrqcd_scaling}. These authors rescale their 
results to the Landau scheme and found improved scaling.
Here we find some indication for a minor improvement, but none
of the schemes is completely successful in removing the apparently large 
${\cal O}(\alpha a)$ errors. A non-perturbative determination of 
the clover coefficients is in order to reduce those errors significantly.
We used the Landau scheme mainly for simplicity.

\begin{figure}[t]
\epsffile{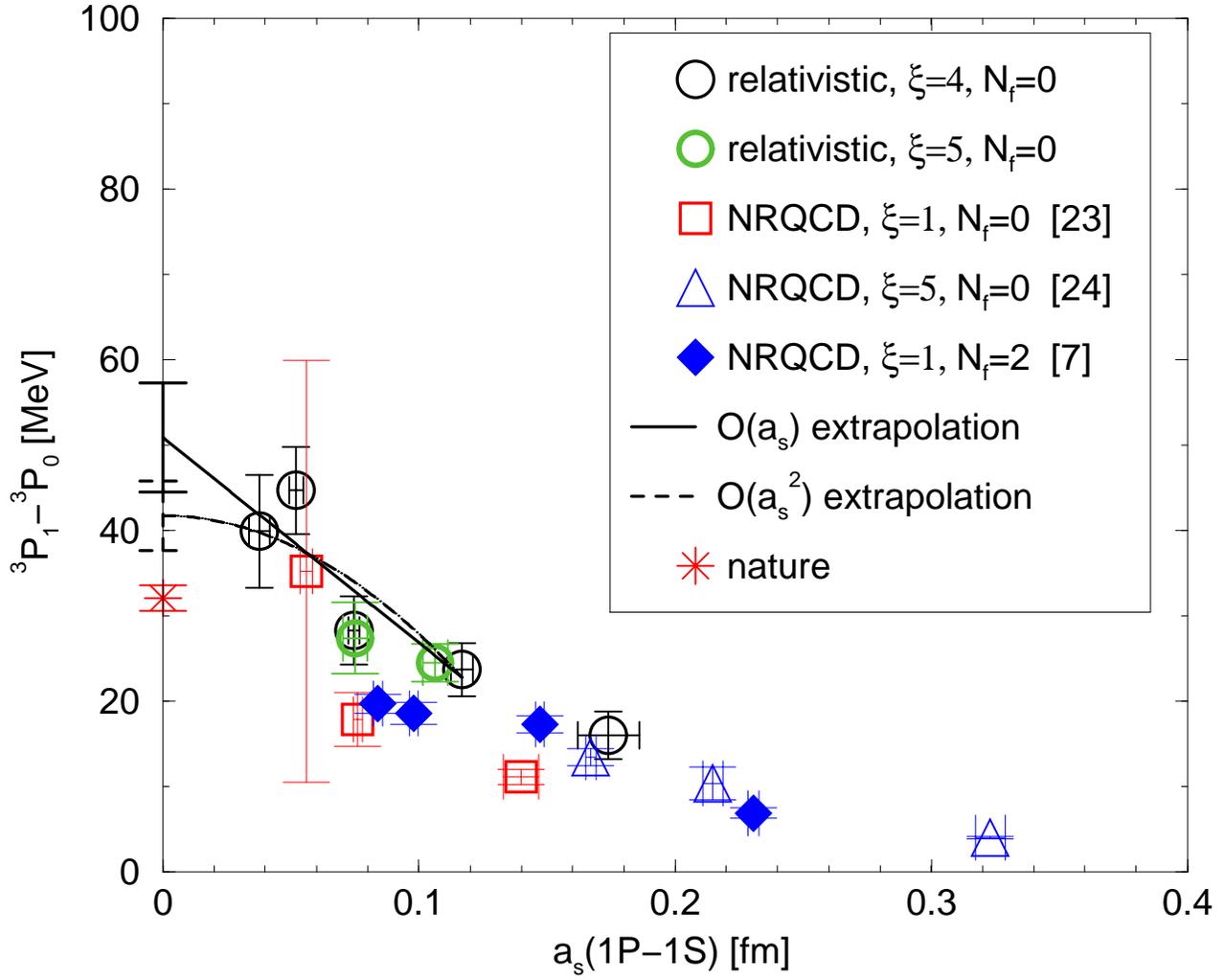}
\caption{Continuum extrapolation of ${^3P}_1-{^3P}_0$. We use the 
same symbols as in Figure \ref{fig:hfs_vs_as} to plot this element
of the fine structure against the spatial lattice spacing
determined from $1P-1S$.}
\label{fig:fs}
\end{figure}

In contrast, the fine-structure is not expected to be as much affected by 
the tadpole prescription which can explain the better agreement of
our estimates for ${^3P}_1-{^3P}_0$ with the NRQCD values in
Figure \ref{fig:fs}.

We interpret the remaining differences between our results and the 
quenched NRQCD values as due to higher order relativistic effects.
It has already been shown in \cite{omv6,spitz} that those corrections 
can be sizable ($>10\%$ for bottomonium at ${\cal O}(mv^6)$). 
However, it is not trivial to extend the non-relativistic approximation to 
ever higher order. Our calculation does not suffer from this 
uncertainty since we are working in a fully relativistic setting.

Above all we are now able to perform a continuum extrapolation
of the hyperfine splitting. 
As shown in Figure \ref{fig:hfs_vs_as} the scaling violations are large 
and we parameterise our lattice results by 1) a linear ansatz, 
2) by a quadratic ansatz and 3) by a linear-plus-quadratic fit in $a_s$. 
All these fits are acceptable with $Q=0.3\ldots 0.9$.
For the final analysis we also decided to omit the results from our coarsest 
lattice where we should expect potentially large lattice spacing corrections 
from higher orders due to $a_t m_q > 1$ and $a_s m_q v >1$.
Including this data, however, does not change our continuum estimates 
significantly. 
If we assume that our action successfully removes all ${\cal O}(a)$ errors, 
we can quote 51.1(3.1) MeV for the quenched hyperfine
splitting in bottomonium. This may be too optimistic and, allowing
for ${\cal O}(\alpha a)$ errors, we find 59(20) MeV from a 
linear-plus-quadratic fit. As can already be seen 
from the large error on the intercept, such a general fit is not very well 
constrained by our data and even the quadratic term is consistent with zero.
For comparison, we find 58.7(5.5) MeV from a simple linear fit.
Notice though, that all these fitting methods result in comparatively 
high values given phenomenological estimates of 30-40 MeV. 
It remains to be seen how unquenching will affect this quantity,
although we expect that it would further increase the hyperfine splitting.
An accurate experimental determination of the bottomonium hyperfine splitting
would be most valuable to judge the reliability of our approach.

For the fine structure, ${^3P}_1-{^3P}_0$, the situation is very similar
as shown in Figure \ref{fig:fs}.
We find a rather large value of 50.9(6.4) MeV from the simple linear
extrapolation in $a_s$ ($Q=0.22$). 
This is $3\sigma$ above the experimental value of 32.1(1.5) MeV. 
The quadratic fit is slightly worse ($Q=0.14$) and 
predicts 41.7(4.1) MeV for the continuum limit. 
As for the hyperfine splitting, a fit involving both terms is 
badly constrained and gives 68(22) MeV with $Q=0.12$.

\begin{figure}[t]
\epsffile{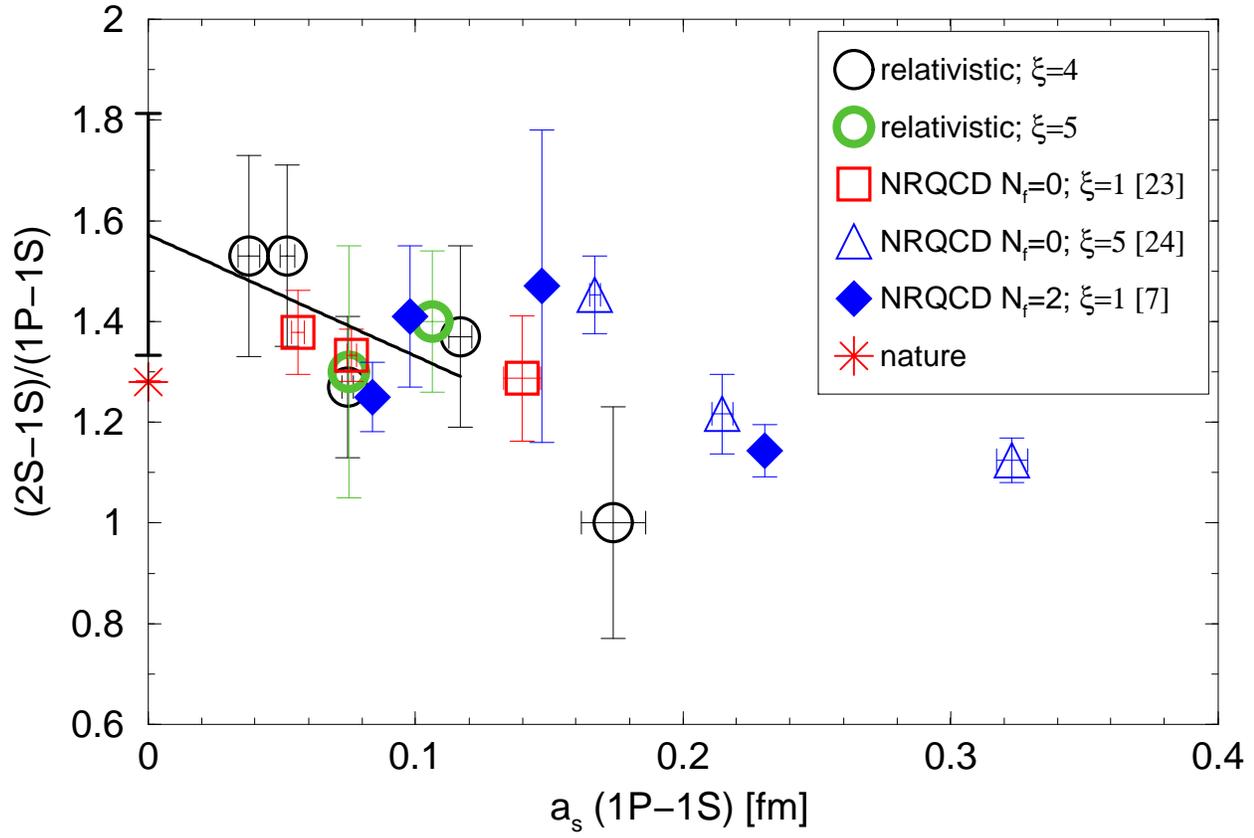}
\caption{Higher excitation in bottomonium, $\Upsilon^\prime$.
Here we plot the conveniently normalised splitting, 
$2S - 1S$, against the spatial lattice spacing. Within the
errors we find a good agreement for all different methods and 
a slightly higher continuum limit than experiment. 
From potential models, we can expect that this ratio will be 
lower in dynamical simulations.}
\label{fig:R2S}
\end{figure}

In Figure \ref{fig:R2S} we also plot the higher excitation, $2S-1S$,
against the spatial lattice spacing. Naturally these results have 
larger errors than spin-splittings and ground states as they rely 
on the control of additional parameters (amplitudes and energies) 
in the fitting process. 
In line with other simulations, we find a continuum result which is 
slightly larger than the experimental value. 
However, it will be important to measure this quantity more accurately 
on bigger volumes as our lattices may not be big enough for those 
larger states. Based on the observed finite volume effects, 
Figure \ref{fig:finite_volume}, we have reason to believe that the 
result from our finest lattice $(\beta,\xi)=(6.5,5)$ could be overestimated. 
Therefore we treat our data for the $2S-1S$ splitting with caution.

\section{Conclusion}
\label{sec:conclusion}
In conclusion, we have demonstrated that also systems containing 
quarks as heavy as bottom can be treated relativistically within 
the framework of {\it anisotropic} lattice QCD. 
Imposing a fine temporal discretisation appears to be a very natural 
way to respect the physical scales in the problem and it should open 
up a reliable alternative to non-relativistic simulations (where 
systematic errors are much harder to control).
In our study of the bottomonium system we found noticeable
deviations of the hyperfine splitting from non-relativistic simulations,
while the fine-structure agrees well with previous results.
Overall we  observed strong scaling violations resulting in a
relatively large value and large errors for the continuum limit. 
This however, is not totally unexpected as we only used a tree-level 
estimate for the clover coefficients and did not tune them non-perturbatively 
to eliminate ${\cal O}(a_s)$ effects completely. 
We believe that further studies on anisotropic lattices will be able
to determine the continuum spectrum of bottomonium with increasing
accuracy.
A non-perturbative determination of the clover coefficients is
highly desirable, but it remains to be seen whether this is feasible
as suggested in \cite{klassen_sf}. Alternatively, simulations
at even finer lattice spacings would be useful to disentangle
remaining ${\cal O}(a_s)$ and ${\cal O}(a_s^2)$ errors.
What remains to be done is the efficient implementation of 
anisotropic sea quarks which is necessary to control the systematic errors 
from the quenched calculation. 
Work in this area is in progress \cite{ludmila}.

Future applications are not restricted to heavy quark masses alone, 
but they may also be extended to heavy-light systems where both 
quarks can now be treated within a uniform approach.

\begin{center}{\bf Acknowledgments}\end{center}
This work was conducted on the QCDSP machines at 
Columbia University and RIKEN-BNL Research Center.
We like to thank N. Christ and R. Mawhinney for
their valuable comments and suggestions. 
TM and XL are supported by the U.S. Department of Energy.

\newpage

\begin{table}
\begin{center}
\begin{tabular}{llllllllllll}
\hline                                                                                                 
$(\beta,\xi)$                   &  (5.7,4) & (5.9,4)    &  (5.9,5)  & (6.1, 4)   &(6.1, 4)   & (6.1, 5)  & (6.3 ,4)  & (6.5 ,4)   \\
$(N_s,N_t)$                     &  (8, 96) & (8, 96)    &  (8, 96)  & (16,96)    &(8,96)     & (16,128)  & (16,128)  & (16,160)   \\
configs                         &  392     &  700       &  780      &  660       & 420       & 370       &  450      & 710         \\
separation                      &  100     &  100       &  400      &  400       & 400       & 400       &  400      & 400         \\
\hline                                                                                                  
$a_t^{-1}$ [GeV]                & 4.54(31) & 6.76(24)   & 9.27(43)  & 10.57(31)  & 9.65(75)  & 12.3(1.2) & 15.15(81) & 20.9(2.2)     \\
$a_s$ [fm]                      & 0.174(12)& 0.1168(42) & 0.1064(49)& 0.0747(22) & 0.0818(64)& 0.0804(77)& 0.0521(28)& 0.0377(40)     \\
\hline                                                                                                 
$\xi_0$                         & 3.04682  & 3.139035   & 3.870249  & 3.210801   & 3.210801  & 3.96234   & 3.268645  & 3.31655      \\
$u_{0s}$                        & 0.758364 & 0.785945   & 0.784067  & 0.800927   & 0.800927  & 0.81015   & 0.826810  & 0.83709      \\
$u_{0t}$                        & 0.984104 & 0.986984   & 0.991782  & 0.988702   & 0.988702  & 0.99279   & 0.990039  & 0.99100      \\
$u_{0t}/u_{0s}$                 & 1.297667 & 1.255793   & 1.26492   & 1.234447   & 1.234447  & 1.22544   & 1.197420  & 1.18386      \\
$\xi/\xi_0$                     & 1.312844 & 1.274277   & 1.2919065 & 1.245795   & 1.245795  & 1.26188   & 1.223749  & 1.20607     \\
$a_tm_q$                        & 1.98     & 1.1200     & 0.8960    & 0.6700     & 0.6700    & 0.58      & 0.4940    & 0.31        \\
($\nu_s, \nu_t$)                & (1, 1.30)& (1, 1.50)  & (1, 1.815)& (1, 1.573) & (1, 1.573)&(1,1.65)   & (1, 1.5)  & (1,1.51)     \\
$c_s$                           & 2.292804 & 2.059794   & 2.074632  & 1.946351   & 1.946351  & 1.88063   & 1.7680    &  1.7054       \\
$c_t$                           & 1.170549 & 1.127661   & 1.1177481 & 1.098368   & 1.098368  & 1.05055   & 1.0154    & 0.9002       \\
$c(0)$                          & 1.044(25)& 1.026(32)  & 0.990(11) &  0.979(7)  & 1.018(14) & 0.994(11) & 1.004(19) & 0.992(20)       \\
\hline                                                                          
${^3}S_1$  [GeV]                & 9.79(66) & 9.53(34)   & 9.57(44)  & 10.40(30)  &  9.49(74) & 10.11(97) & 12.60(67) & 13.04(14)    \\
1 ${^3}S_1-{^1}S_0$  [MeV]      & 36.0(3.5)& 37.2(2.6)  & 37.5(2.9) & 44.8(2.0)  & 33.8(3.1) & 37.0(5.7) & 49.9(4.9) & 51.0(7.9)    \\
2 ${^3}S_1-{^1}S_0$  [MeV]      & 155(273) & 56(57)     & 37(12)    & 40(11)     & 57(18)    &   2(58)   &  48(41)   & 57(29)       \\
${^1}P_1-{^3}P_0$  [MeV]        & 16.2(4.7)& 20.6(4.5)  & 25.2(3.2) & 39.5(5.7)  & 28.7(6.6) & 30.7(5.5) & 54.0(7.7) & 46.6(8.7)          \\
${^1}P_1-{^3}P_1$  [MeV]        & -1.4(2.3)&-1.3(1.1)   & 4.6(1.3)  & 8.0(3.3)   &  7.9(4.9) &  3.7(2.0) &  8.8(3.5) &  8.4(3.8)           \\
${^3}P_1-{^3}P_0$  [MeV]        & 16.0(2.8)& 23.7(3.1)  & 24.5(2.2) & 28.3(4.0)  & 28.0(5.7) & 27.4(4.2) &  44.7(5.1)& 39.9(6.6)            \\
\hline                                                                          
$R_{2S}$                        & 1.00(23) & 1.37(18)   & 1.40(14)  & 1.27(14)   & 1.38(17)  & 1.30(25)  &  1.53(18) & 1.53(20)       \\
$R_{3S}$                        & 1.99(50)*& 2.35(29)*  & 3.13(32)* & 2.27(34)*  & 3.6(1.3)* & 3.05(57)* &  3.25(51)*& 5.17(76)*     \\
$R_{2P}$                        & 2.21(48)*& 2.32(30)   & 2.26(26)  & 2.40(13)   & 2.26(30)  & 1.98(61)  &  2.56(49) & 2.14(34)     \\
\end{tabular}
\caption{Simulation Parameters and Results. We label our runs by $(\beta, \xi)$,
where $\beta$ is chosen to vary the spatial lattice spacing
and $\xi$ determines the appropriate bare anisotropy $\xi_0$, see Eq. \ref{eq:renorm_aniso}.
The temporal lattice spacing, $a_t$, is determined from ${^1}P_1-{^3}S_1$.
Ground state and higher excitations are from correlated 3-cosh fits to 
all channels. For the excitations we quote their normalized results
$R_X \equiv (X-1S)/(1P-1S)$.
An asterisk denotes results which have not been checked 
by fitting even higher excitations than those listed.
For example, we did not perform 4-cosh fits to obtain
better estimates for $3S$.
All spin-splittings have been rescaled to the physical bottomonium  
(${^3S}_1=9.46$ GeV). }
\label{tab:results}
\end{center}
\end{table}

\begin{table}
\begin{center}
\begin{tabular}{llllllllllll}
\hline                                           
$(\beta,\xi)$                   & (6.1, 4)   & (6.3 ,4)   \\
configs                         &  330       &  360        \\
separation                      &  400       &  400        \\
\hline                                       		 
${^3}P_2-{^1}P_1$  [MeV]        & 24.4(4.1)  & 20.6(6.9)              \\
${^3}P_2-{^3}P_1$  [MeV]        & 39(14)     & 27(11)        \\
$2^{--}-2^{-+}$    [MeV]        & --         & 31.0(6.9)      \\
\hline                                       		 
$R_{2^{-+}}$                    &  --        &  1.63(11)             \\
$R_{2^{--}}$                    & 1.81(26)   &  1.67(11)           \\
\hline                                       		 
$R_{1^{-+}}$                    & 3.41(43)   &  4.25(34)           \\
$R_{0^{+-}}$                    & 4.41(85)   &  3.37(72)       \\
$R_{2^{+-}}$                    &  --        &  5.93(45)          \\
\end{tabular}
\caption{Results from extended meson operators. For the two
lattices shown all run parameters are exactly as in Table \ref{tab:results}
and $R_X$ is the normalized excitation above the ground state:
$R_X \equiv (X-1S)/(1P-1S)$.
However, here we employ also the extended operators of 
Eq. (\ref{eq:operators}) which give us access to ${^3P}_2$, states with
$L=2$ and the exotic hybrid candidates $(0^{+-}, 1^{-+}, 2^{+-})$.
For P-states ($L=1$) we use a single symmetric lattice derivative 
to create extended operators and for D-states we take
the anticommutator $\{\Delta_i, \Delta_j\}$,  ($L=2$).
The commutator $[\Delta_i, \Delta_j]$ amounts to an
insertion of a colour-magnetic field into the quark bilinear
and gives rise to hybrid states. Different Dirac matrices
result in different $J^{PC}$ assignments.}
\label{tab:exotics}
\end{center}
\end{table}

\begin{table}
\begin{center}
\begin{tabular}{lllllllll}
\hline                                                  
$(\nu_s, \nu_t)$                &  (1., 1.815)              &  (1.,1.650)             \\
$(c_s,c_t)$                     & (2.074632, 1.1177481)     &  (2.074632, 1.090685 )  \\
\hline                                                  
configs                         &  780                      &  500       \\
\hline                                                  
$c(0)$                          & 0.990(11)                 & 1.067(18)    \\
$a_t^{-1}$ [GeV]                & 9.27(43)                  & 8.20(34)   \\
\hline                                                  
${^3}S_1$  [GeV]                & 9.57(44)                  & 9.05(38)    \\
${^3}S_1-{^1}S_0$  [MeV]        & 37.5(2.9)                 & 34.4(2.6)   \\
${^1}P_1-{^3}P_0$  [MeV]        & 25.2(3.2)                 & 24.3(5.7)    \\
${^1}P_1-{^3}P_1$  [MeV]        &  4.6(1.3)                 &  4.6(1.4)   \\
${^3}P_1-{^3}P_0$  [MeV]        & 24.5(2.2)                 & 24.6(2.5)    \\
$R_{2S}$                        & 1.40(14)                  & 1.324(81)    \\
$R_{2P}$                        & 2.26(26)                  & 2.54(23)   \\
\end{tabular}
\caption{Change in $\nu_t$. In this table we comapre our results from $(\beta,\xi)=(5.9,5)$ 
when the optimal $\nu_t=1.815$  is changed arbitrarily by 10\%. 
This entails a change in $c_t$. All other run parameters
are as in column 4 of Tab. \ref{tab:results}. The corresponding changes in the spectrum
as listed above and the lattice numbers are given in Fig. \ref{fig:nut_dependence}.}
\label{tab:nut}
\end{center}
\end{table}

\begin{table}
\begin{center}
\begin{tabular}{lllllllll}
tadpole scheme                  & $u_{0L}$   & $u_{0P}$   & $u_{0L}$  & $u_{0P}$    \\
\hline                                                                            
$(\beta,\xi)$                   & (5.9,4)    & (5.9,4)    & (6.1, 4)  & (6.1, 4)  \\
$(N_s,N_t)$                     & (8, 64)    & (8, 64)    & (16, 96)  & (16, 96)   \\
configs                         &  600       &  600       &  660      &  130       \\
\hline                                                                  
$a_t^{-1}$ [GeV]                & 6.57(43)   &  6.40(53)  & 10.57(31) &  10.6(1.1)  \\
$a_s$ [fm]                      & 0.1202(80) &  0.123(10) & 0.0747(22)& 0.0745(77)            \\
\hline                                                                  
$\xi_0$                         & 3.139035   & 3.139035   & 3.210801  &  3.210801   \\
$u_{0s}$                        & 0.785945   & 0.810698   & 0.800927  &  0.822785   \\
$u_{0t}$                        & 0.986984   & 0.949731   & 0.988702  &  0.952954   \\
$a_sm_q$                        & 1.1200     & 1.1200     & 0.6700    &  0.877      \\
$(\nu_s,\nu_t)$                 & (1.,1.50)  & (1.,1.50)  & (1.,1.573)& (1.,1.573)  \\
$c_s$                           & 2.059794   & 1.876818   & 1.946351  &  1.795316    \\
$c_t$                           & 1.127661   & 1.101421   & 1.098368  &  1.079827   \\
$c(0)$                          & 1.014(15)  & 1.023(14)  & 0.979(7)  &   --        \\
\hline                                                                  
${^3}S_1$  [GeV]                & 9.25(61)   & 9.04(75)   & 10.40(30) &  12.3(1.3)  \\
${^3}S_1-{^1}S_0$  [MeV]        & 35.1(3.9)  & 30.6(3.6)  & 44.8(2.0) &  43.7(6.6)  \\
${^1}P_1-{^3}P_0$  [MeV]        & 26.0(3.9)  & 25.0(4.0)  & 39.5(5.7) &  41(14)      \\
${^1}P_1-{^3}P_1$  [MeV]        &  1.3(1.7)  &  1.7(2.3)  &  8.0(3.3) &  9.6(8.2)     \\
${^3}P_1-{^3}P_0$  [MeV]        & 23.7(2.7)  & 22.3(2.8)  & 28.3(4.0) & 30.8(9.4)    \\
\hline
$R_{2S}$                        & 1.45(13)   & 1.41(13)   & 1.27(14)  & 1.37(23)*      \\
$R_{2P}$                        & 2.22(49)   & 1.81(28)   & 2.40(13)  & 3.14(97)*     \\
\end{tabular}
\caption{Here we compare our results from different tadpole descriptions at two different
lattices spacings. We distinguish between the Landau scheme, $u_{0L}$, as defined in Section 
\ref{sec:actions} and the plaquette scheme, $u_{0P}$, for which  
$3 u_\mu= \langle {\rm tr}~P_{i \mu} \rangle^{1/4}$.
While there is some indication that the hyperfine splitting increases when $u_0$ is decreased
at finite lattice spacing, there is no clear improvement of scaling. See discussion in main text.}
\label{tab:tadpole}
\end{center}
\end{table}

% FIGURES

\newpage

\end{document}